\documentclass[runningheads]{llncs}

\usepackage{graphicx}
\usepackage{color}
\usepackage{subcaption}
\usepackage{amsmath}
\usepackage{booktabs}
\usepackage[table,xcdraw]{xcolor}
\usepackage[breaklinks=true]{hyperref}

\begin{document}

\title{AGE-US: automated gestational age estimation based on fetal ultrasound images}

\titlerunning{AGE-US}

\author{César Díaz-Parga\inst{1,2} 
\and Marta Nuñez-Garcia \inst{1,2}
\and Maria J. Carreira \inst{1,2}
\and Gabriel Bernardino \inst{3} 
\and Nicolás Vila-Blanco \inst{1,2} 
}

\authorrunning{C. Díaz-Parga et al.}

\institute{Centro Singular de Investigación en Tecnoloxías Intelixentes (CiTIUS), Universidade de Santiago de Compostela, 15782 Santiago de Compostela, Spain \and Departamento de Electrónica e Computación, Escola Técnica Superior de Enxeñaría, Universidade de Santiago de Compostela, 15782 Santiago de Compostela, Spain \and BCN-MedTech, Universitat Pompeu Fabra, 08018 Barcelona, Spain}

\maketitle              
\begin{abstract}
Being born small carries significant health risks, including increased neonatal mortality and a higher likelihood of future cardiac diseases. Accurate estimation of gestational age is critical for monitoring fetal growth, but traditional methods, such as estimation based on the last menstrual period, are in some situations difficult to obtain. While ultrasound-based approaches offer greater reliability, they rely on manual measurements that introduce variability. This study presents an interpretable deep learning-based method for automated gestational age calculation, leveraging a novel segmentation architecture and distance maps to overcome dataset limitations and the scarcity of segmentation masks. Our approach achieves performance comparable to state-of-the-art models while reducing complexity, making it particularly suitable for resource-constrained settings and with limited annotated data. Furthermore, our results demonstrate that the use of distance maps is particularly suitable for estimating femur endpoints.

\end{abstract}

\section{Introduction}
Being born small is a global health challenge with serious consequences including perinatal mortality and the risk of developing future cardiac diseases. In recognition of its importance, a recent series of articles in The Lancet introduced the term Small and Vulnerable Newborns (SVN), which encompasses all causes of being born small~\cite{ashorn2020lancet}. This 
condition affects an estimated 36.5 million births each year and accounts for nearly 1.4 million deaths (half of all neonatal mortality) worldwide~\cite{lawn2023small}.

To address this issue and ensure optimal maternal-fetal health outcomes, fetal growth must be effectively monitored. Ultrasound (US) imaging is used to estimate the size of the foetus, which is then compared to normality ranges to determine whether the size is correct or small for the gestational age (GA).
The accurate estimation of GA is therefore essential for assessing fetal growth and determining the appropriate timing of medical interventions. 
GA is typically estimated from the date of the last menstrual period (LMP). However, it relies on the recall of an exact date, which can be challenging in low- and middle-income countries where screening is typically performed only in the third trimester~\cite{Bota2021Metabolic}, and for women with irregular menstrual cycles~\cite{Papageorghiou2016Ultrasound}. 
When LMP is not available, US imaging is used as an alternative method for the estimation of GA.
The standard procedure involves the acquisition of specific fetal anatomical planes and the measurement of biometrics such as the head circumference, the biparietal diameter, the abdominal circumference, and the femur length. In clinical practice, these measurements are performed manually, which is time-consuming and operator-dependent (inter-observer variability may range from 4.9\% to 11\% according to~\cite{sarris2012intra}).
Automation of fetal biometrics is a promising solution to reduce intra- and inter-observer variability, minimise assessment time, and reduce costs~\cite{kurjak2017donald}. This is particularly important in regions with limited access to US scanners and experienced sonographers. 

In medical imaging, machine learning techniques have demonstrated efficacy in diagnosis and outcome prediction, even outperforming human efficiency in tasks such as segmentation and automated quantification~\cite{BarraganMontero2021Artificial}. 
Specifically, deep learning models may be useful for fetal biometric measurements and fetal growth assessment~\cite{seval2023current}.
But despite its potential, the application of deep learning techniques to fetal biometric estimation remains challenging. 
A major limitation is the relatively small size of medical datasets compared to other domains. This is mainly due to the lack of standardised storage, annotation, and acquisition protocols in clinical routine. 
The challenge is even greater for fetal growth abnormalities, which have lower incidence rates, further limiting the amount of data available. 
As a result, models trained on small datasets are prone to overfitting, reducing their ability to generalise across different devices and clinical settings. Fine-tuning techniques offer a potential solution by exploiting larger public datasets and transferring the learned knowledge, thus achieving better results than training models solely on small individual datasets.

This work proposes AGE-US\footnote{Source code and model weights: \url{https://gitlab.citius.gal/cesar.diaz.parga/age-us}}, a method for automated GA estimation based on US images of the head, abdomen, and femur planes acquired in the third trimester of pregnancy.
Our method uses transfer learning and distance maps for structures where segmentation masks are scarce, making it suitable for smaller image datasets. 
The main contributions of this study include (1) a single-encoder dual-decoder U-Net architecture that reduces the complexity of the segmentation model without statistically significant performance degradation, and (2) a method for addressing the unavailability of femur segmentation masks by incorporating distance maps. Finally, to provide a more interpretable alternative to end-to-end GA estimation methods~\cite{Dan2023Deep}, AGE-US integrates these contributions in a fully automated pipeline that mimics standard clinical practice.

\section{Related work}

Several approaches utilise architectures such as the U-Net~\cite{Ronnenber2015MICCAI} to segment the head and then estimate the head circumference (HC), e.g. Kim \textit{et al.} who refine the head boundaries based on the location of specific brain regions~\cite{Kim2019Physiological}.
Other methods employ fully connected architectures and ellipse fitting over segmentation contours to compute HC and biparietal diameter (BPD)~\cite{Sinclair2018EMBC}. In addition, DAG V-Net~\cite{zeng2021fetal} incorporates attention mechanisms into a V-Net to improve the accuracy of HC measurements refine the head boundaries based on the location of specific brain regions~\cite{Kim2019Physiological}.
Other methods employ fully connected architectures and ellipse fitting over segmentation contours to compute HC and biparietal diameter (BPD)~\cite{Sinclair2018EMBC}.
In addition, DAG V-Net~\cite{zeng2021fetal} incorporates attention mechanisms into a V-Net to improve the accuracy of HC measurements.

With respect to the estimation of the abdominal circumference (AC), Jang \textit{et al.} proposed to detect key areas in the abdominal plane and apply the Hough transform to obtain accurate measurements~\cite{Jang2018Ultrasonic}.
However, this method suffers from a significant limitation: it relies on segmenting increasingly complex areas to accurately locate the AC, leading to a drastic decrease in performance in the presence of image shadows over these areas. 
AC estimation is often integrated into broader methods that segment multiple structures. 
For example, FUVAI~\cite{plotka2022deep} is a CNN-based U-Net architecture that combines plane classification and segmentation to compute HC, BPD, AC, and femur length (FL) from video sequences.
Other methods apply a similar approach to 2D images, using a U-Net to segment all structures in different planes, followed by ellipse fitting for the abdominal plane and rectangle fitting for the femur plane~\cite{Qazi2023MICAD}.
However, all these methods rely on a large amount of labelled data across multiple planes, which are generally not publicly available.

Once the biometric measures are obtained, they are used to estimate the GA. Various equations have been proposed in this regard, with Hadlock’s equation~\cite{Hadlock1984GA} being the most widely used in its 4-parameter version.
Another alternative is Skupski’s equation~\cite{skupski2017estimating}, which also utilises 4 parameters. 

Fully automated methods such as DeepGA~\cite{Dan2023Deep} combine segmentation and regression modules to compute GA directly from images without relying on explicit biometric measurements. While this improves automation, it requires large amounts of training data and lacks interpretability in how predictions are made.
Publicly available fetal biometric data are scarce and mainly focused only on the head plane (e.g. the HC18 dataset~\cite{Van_den_Heuvel2018HC18}). The ACOUSLIC-AI challenge dataset~\cite{MICCAI2024ACOUSLIC} contains images of the abdominal plane, but only few of them are annotated (only those belonging to the optimal plane).
To the best of our knowledge, no publicly available dataset includes femur masks or femur measurements.

Given that the available data is insufficient to adequately train most methods in the literature, our approach exploits the greater availability of labelled head data to improve training for the abdominal plane.
This is achieved by first initialising the model with a large public annotated head set and a small set of custom-labelled abdominal data, and then fine-tuning it on the specific training dataset. 
Additionally, our method mitigates the scarcity of femur masks by exploiting weak annotations over the femur and focusing on endpoint detection rather than full bone segmentation.

\section{Methodology}

\begin{figure}[tp]
     \centering
     \includegraphics[width=0.99\linewidth]{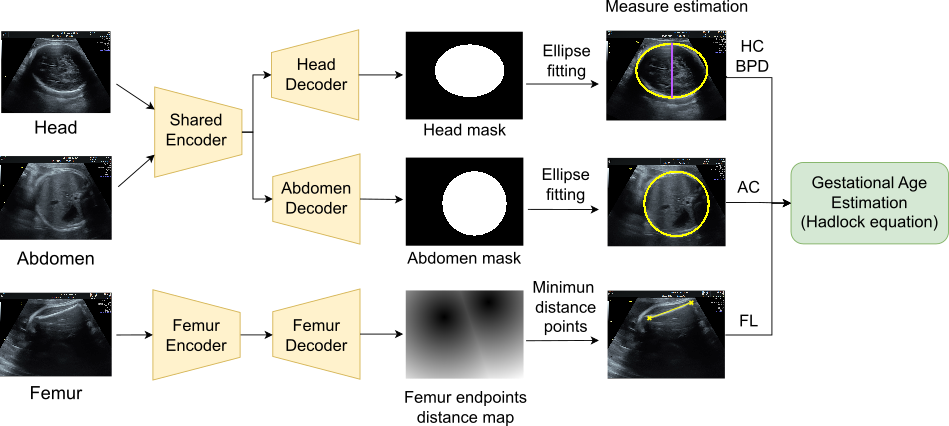}
     \caption{AGE-US workflow. First, abdomen and head masks are extracted using a single-encoder, dual-decoder U-Net, and a distance map of the femur endpoints is obtained using a conventional U-Net.
     Second, an ellipse is fitted to both the head and abdomen masks to calculate the corresponding circumferences (HC, AC), and the biparietal diameter (BPD), which is approximated by the minor axis of the head ellipse.
     In addition, the estimated femur endpoint location is extracted from the distance map, and its length is calculated.
     Thirdly, the 4 measurements are used in Hadlock's equation to compute the gestational age.}
     \label{Fig:Sistem_diagram}
\end{figure}

Fig.~\ref{Fig:Sistem_diagram} illustrates the AGE-US framework, which automates the measurement process on fetal US images, requiring only 3 specific anatomical planes: head, abdomen, and femur.
It starts with a single-encoder, dual-decoder U-Net architecture designed to efficiently combine the head and abdomen regions of interest. It constitutes a more efficient approach to the calculation of HC, BPD, and AC than the use of two fully independent U-Nets. For the femur, we introduce a distance map-based prediction model that detects the endpoints of the femur and calculates its length as the distance between them. Finally, we use this extracted biometric measurements (i.e. HC, BPD, AC, and FL) and the Hadlock 4-parameter equation to compute the GA in weeks.

\subsection{Head and abdomen segmentation}

As mentioned above, there are 2 major challenges in training models for this task: data scarcity (very few fetal biometric datasets are available) and data imbalance (they only contain annotations for the head plane).
To address this, we present a novel U-Net-based architecture that leverages knowledge learned from the head plane to improve abdominal segmentation, taking advantage of the structural similarity between these two anatomical regions of interest.  
Specifically, we propose to merge the encoders of 2 segmentation models into a common encoder, while maintaining 2 specialised decoding branches. 
By using a single encoder, we reduce the total number of parameters by 25\%.

Finally, the head mask is fitted with an ellipse whose perimeter corresponds to HC, and whose minor axis corresponds to BPD. Similarly, the abdominal mask is fitted with an ellipse to derive the AC.

\subsection{Femur localisation}

The calculation of the length of the femur is particularly challenging due to the aforementioned lack of annotated data. 
To our knowledge, no publicly available dataset includes femur segmentation masks.
Therefore, alternative strategies are required to exploit the available data, which typically consists of weak annotations (i.e., the location of the 2 femur endpoints).

\begin{figure}
     \centering
     \includegraphics[width=\linewidth]{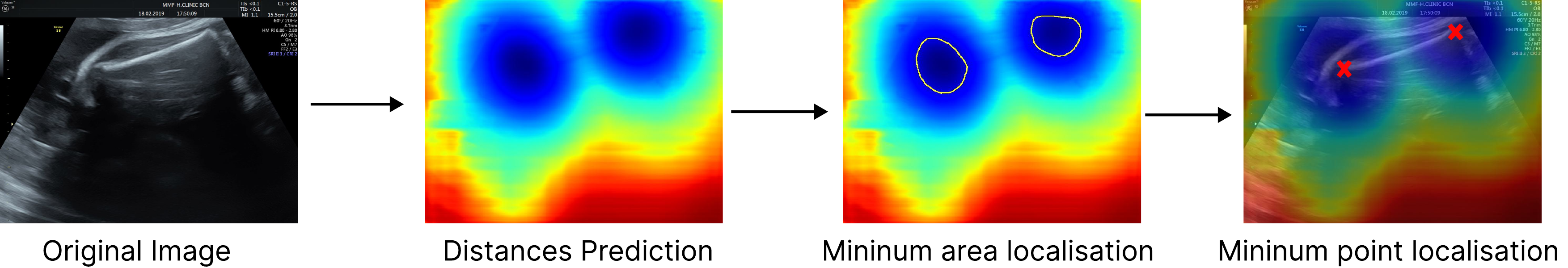}
     \caption{Representation of the method used to locate the femur endpoints (red crosses). 
     Distance values are visualised using a colourmap for clarity, with blue representing the lowest (proximal) values and red representing the highest (distant) values.}
     \label{Fig:point_localization}
\end{figure}

We trained a U-Net-based neural network to predict a distance map encoding pixel-wise distances to the femur endpoints. These maps, normalised between 0 and 1, serve as training targets.
Then, endpoint localisation follows a two-step process, illustrated in Fig.~\ref{Fig:point_localization}: first, the two central regions containing the minima are segmented by using the Watershed algorithm; then, the femur endpoints are extracted as the 2 points with lower values within each region.
Finally, we calculate the FL as the Euclidean distance between the 2 points.

\subsection{Gestational Age Estimation}

Once the necessary biometric measurements have been extracted from each anatomical plane, we calculate the GA. 
For this purpose, we utilise the Hadlock formula, one of the most widely used in clinical practice. 
Specifically, we employ the 4-parameter Hadlock equation as it provides the lowest estimation error among the variants with fewer parameters~\cite{Hadlock1984GA}.
The equation is defined as follows:  

\begin{equation}
    GA(weeks) = 10.85 + (0.060 \cdot HC \cdot FL) + 0.670 \cdot BPD + 0.1680 \cdot AC
\end{equation}  
where HC is the head circumference, BPD is the biparietal diameter, AC is the abdominal circumference and FL is the femur length. 

\section{Experiments and Results}

The experiments were conducted using the IMPACT (Improving Mothers for a Better Prenatal Care Trial) dataset.
It is related to a randomised controlled clinical trial (ClinicalTrials.gov Identifier: NCT03166332) conducted at the Hospital Clínic de Barcelona (Spain) from 2017 to 2020. 
The study included 1,221 high risk pregnancies~\cite{crovetto2021effects}. 
We selected a subset of data based on 2 criteria: the availability of annotated planes for the head, abdomen, and femur; and the presence of at least 1 image for each plane.
As a result, the final number of studies included was 114. The available manually annotated data were head and abdomen segmentation masks, and femur endpoints.
Table~\ref{tab:dataset_values} shows the biometric values based on manual measurements, and the estimation of the GA using the Hadlock equation. 
The GA ranges from 21.6 to 36.3 weeks, with a median of 33.18 weeks.

\begin{table}[]
\setlength\tabcolsep{1em}
\centering
\caption{
Biometric and gestational age values based on manual measurements in the IMPACT dataset (subset of 114 cases).
HC: Head Circumference; BDP: Biparietal Diameter; AC: Abdominal Circumference; FL: Femur length; and GA: Gestational Age.}
\label{tab:dataset_values}
\begin{tabular}{@{}lll@{}}
\cmidrule[1pt](l{0pt}r{0pt}){2-3}
 & \multicolumn{1}{c}{\textbf{Median {[}IQR{]}}} & \multicolumn{1}{c}{\textbf{{[}Min, Max{]}}} \\ \midrule

HC (cm) & \multicolumn{1}{c}{30.514 {[}29.413, 31.485{]}} & \multicolumn{1}{c}{{[}19.676, 34.267{]}}                              \\
BPD (cm) & \multicolumn{1}{c}{8.644 {[}8.319, 8.919{]}} & \multicolumn{1}{c}{{[}5.315, 11.119{]}}                                \\
AC (cm)  & \multicolumn{1}{c}{30.011 {[}28.886,31.201{]}}                            & \multicolumn{1}{c}{{[}12.332, 34.678{]}}                                \\
FL (cm)  & \multicolumn{1}{c}{6.232 {[}6.045,6.502{]}}                            & \multicolumn{1}{c}{{[}4.299, 6.986{]}}                                \\ \midrule
 GA (weeks)  & \multicolumn{1}{c}{33.177 {[}32.345, 33.894{]} }   &   \multicolumn{1}{c}{ {[}21.560, 36.306{]}}                                                                                                                      \\ \bottomrule
\end{tabular}
\end{table}

We used a pre-trained model to improve the segmentation performance.
This model was trained jointly on 2 datasets: the first one comprises 2,873 head images from the head plane, the majority of which belong to the trans-thalamic plane, obtained from the Large-scale Annotation Dataset for Fetal Head Biometry in Ultrasound Images~\cite{ALZUBAIDI2023109708}; the second one contains 420 manually annotated abdominal images from the Fetal\_Planes\_DB dataset~\cite{burgos2020Fetal}.
 
We selected these images because the acquisition devices were very similar, minimising variability in the input data.

The network architecture is derived from the original U-Net model~\cite{Ronnenber2015MICCAI}, employing an encoder-decoder framework with skip connections. The encoder path features 4 downsampling blocks, where each block comprises a MaxPooling operation (2×2 kernel, stride 2) followed by two 3×3 convolutional layers (stride 1, padding 1).
Instance Normalization and a LeakyReLU activation function are applied after each of these convolutional layers.
For the decoder path, which is designed to generate outputs for 2 distinct structures, 8 upsampling blocks are utilized, arranged as 2 parallel pathways of 4 blocks each.
Each upsampling block consists of a Transposed Convolution (e.g., with a 2×2 kernel and stride 2) for spatial upsampling, followed by concatenation of its output with corresponding skip connection features from the encoder.
Subsequently, 2 further 3×3 convolutional layers (stride 1, padding 1) are applied; similar to the encoder, these 2 decoder convolutions are each followed by Instance Normalization and a LeakyReLU activation function. Following the same architectural principles and configurations, the femur model employs a single branch decoder with only 4 upsampling blocks. 

Model pre-training and finetuning were performed using the Adam optimiser with an initial learning rate of 1e-3, using as a loss function the equally weighted sum of Dice Loss and Cross-Entropy Loss. 
We used the entire segmentation datasets for pre-training. For fine-tuning, we employed a subset (75\%) of the IMPACT available data. The intensity of all images and masks was normalised to the range [0,1]. 
Finally, we rescaled all images to a uniform size of 256x256 and no additional data augmentation operations were applied.
For both cases, 10\% of the training data was reserved for validation, which was performed every 2 epochs. We trained the segmentation models for 1,000 epochs and fine-tuned them for 100 epochs on IMPACT, selecting the best model in validation based on the Dice score. 
On the other hand, we used only IMPACT data to train the femur model, training it from scratch for 1,000 epochs. We used the lowest validation loss for the femur localisation model to select the best model. 
Then, we used the remaining 25\% IMPACT studies to assess model performance, ensuring that images from the same study did not appear both in the training and test sets. 
All experiments were conducted with a batch size of 10 on an NVIDIA A100 GPU.

Regarding the performance of the head and abdomen segmentation models, we compared the proposed architecture to a model with separate encoder-decoder blocks for each anatomical region. 
We evaluated the performance (see Table~\ref{tab:seg_abdomen} and Table~\ref{tab:seg_brain}) using 2 segmentation metrics: the Dice score for overlap, and the Hausdorff distance for boundary-distance accuracy. 
For each case, we reported the median, interquartile range (IQR), and the worst 5\% of cases. For the Dice score, this corresponds to values below the 5th percentile, while for the Hausdorff distance, it corresponds to values above the 95th percentile.
We conducted hypothesis testing to determine statistical significance. 
First, we applied the Kolmogorov-Smirnov normality test. Since normality was not satisfied, we used the Wilcoxon signed-rank test as a non-parametric alternative to the t-test, with a significance threshold of $p=0.05$.

\begin{figure}[ht]
    \centering
    \begin{subfigure}{0.45\textwidth}
        \centering
        \includegraphics[width=0.85\textwidth]{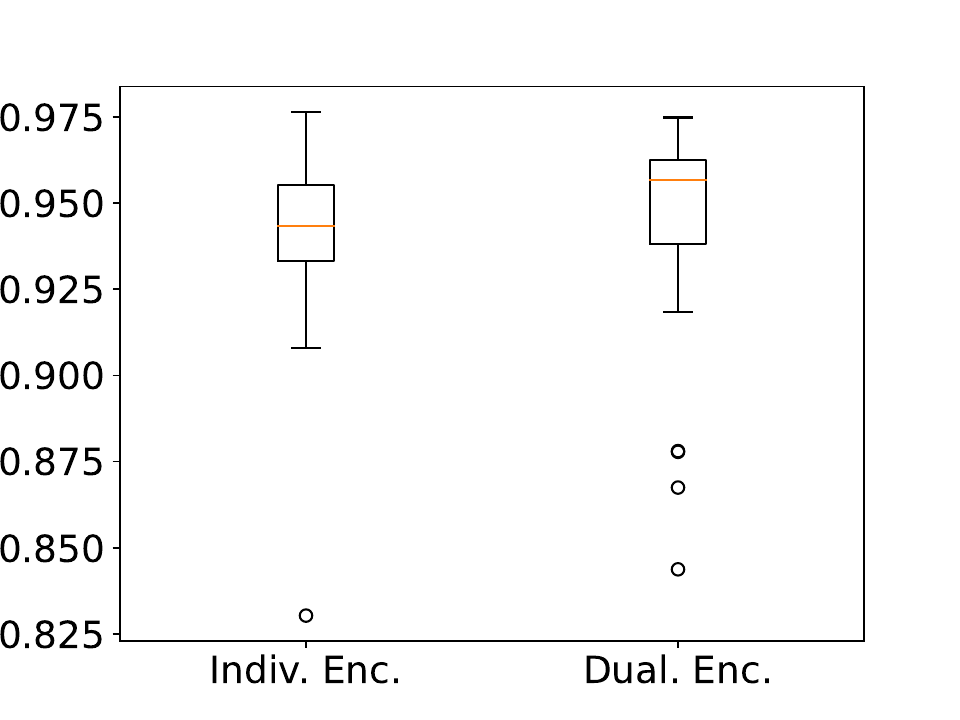} 
        \caption{Dice score for abd. plane.}
        \label{fig:boxplot-dice-abd}
    \end{subfigure} \hfill
    \begin{subfigure}{0.45\textwidth}
        \centering
        \includegraphics[width=0.85\textwidth]{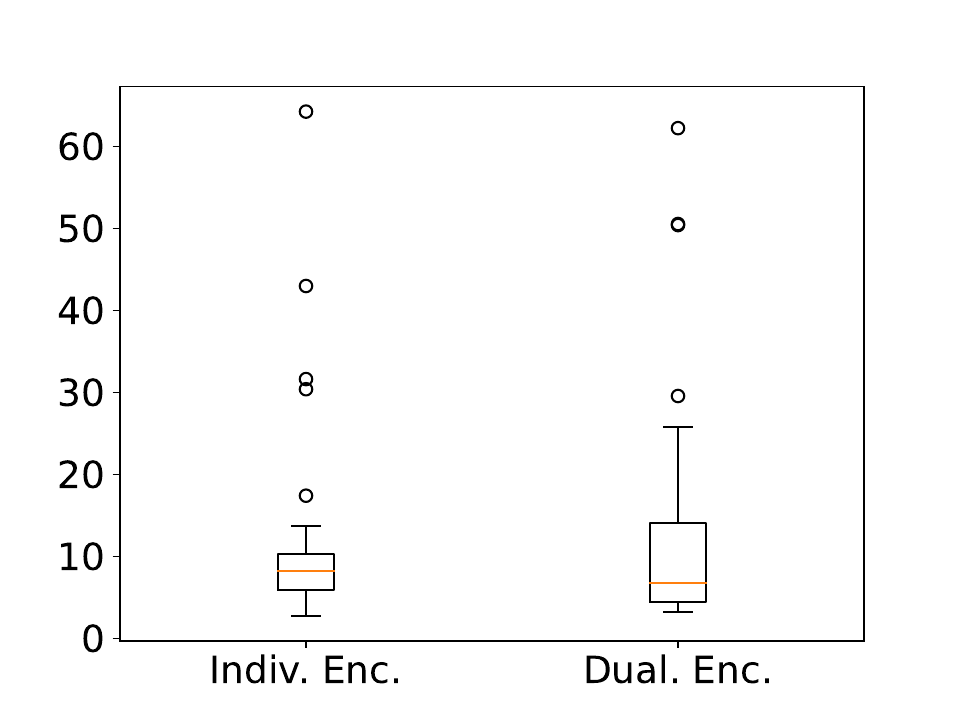} 
        \caption{Hausdorff distance for abd. plane.}
        \label{fig:boxplot-hausdorff-abd}
    \end{subfigure} \hfill
    \begin{subfigure}{0.45\textwidth}
        \centering
        \includegraphics[width=0.85\textwidth]{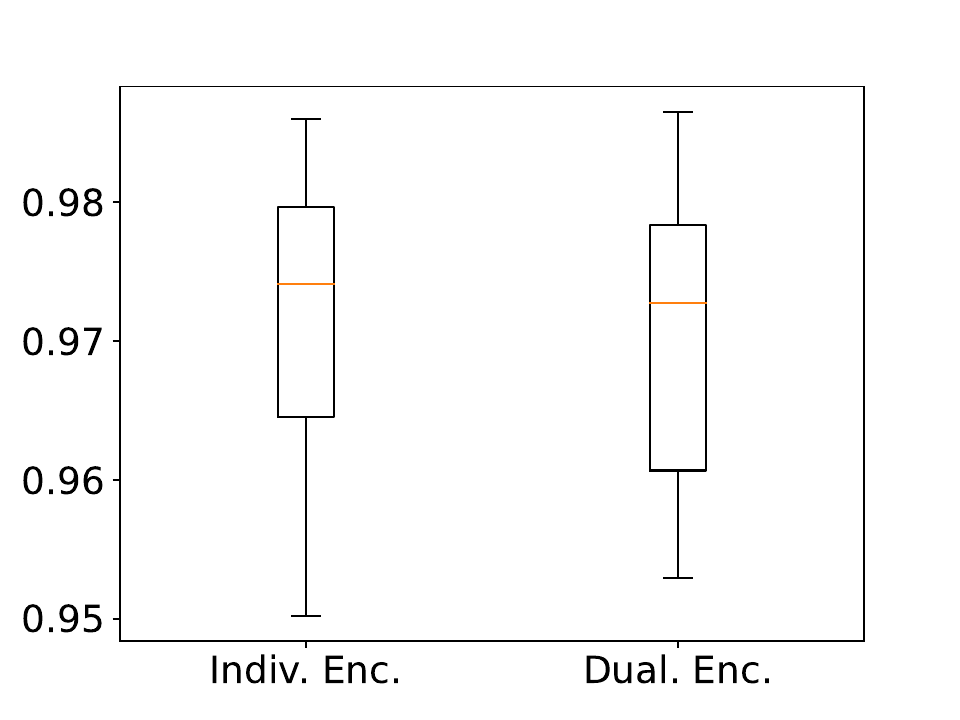} 
        \caption{Dice score for head plane.}
        \label{fig:boxplot-dice-head}
    \end{subfigure} \hfill
    \begin{subfigure}{0.45\textwidth}
        \centering
        \includegraphics[width=0.85\textwidth]{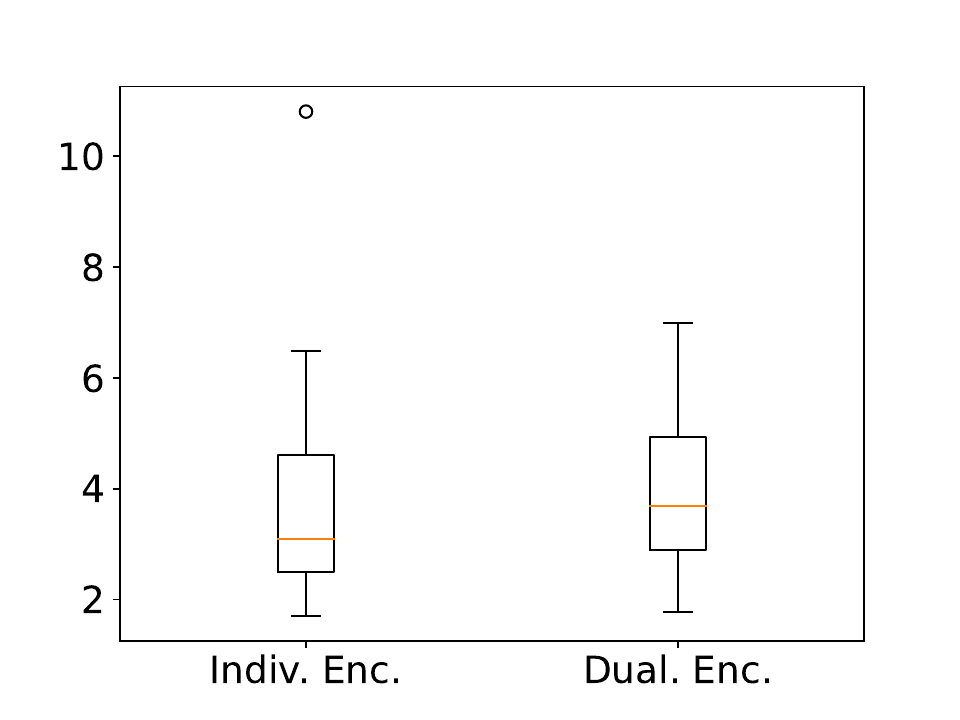} 
        \caption{Hausdorff distance for head plane.}
        \label{fig:boxplot-hausdorff-head}
    \end{subfigure}
    
    \caption{Abdomen and head segmentation results.}
    \label{fig:boxplots-figure}
\end{figure}

With respect to abdominal segmentation, our approach achieves a higher median Dice score (0.956 vs 0.943), and a decreased median Hausdorff distance (6.736 mm vs 8.210 mm) as compared to the individual encoder approach (see Fig.~\ref{fig:boxplot-dice-abd} and~\ref{fig:boxplot-hausdorff-abd}, and Table~\ref{tab:seg_abdomen}). 
The statistical test showed no significant differences between the 2 approaches for both metrics, proving that our model, despite having fewer parameters, achieved performance comparable to a specialised model.
Regarding head segmentation, Fig.~\ref{fig:boxplot-dice-head} and~\ref{fig:boxplot-hausdorff-head}, and Table~\ref{tab:seg_brain}, illustrate a slight (not statistically significant) decrease in the Dice score (0.973 vs 0.974), and a statistically significant increase in the median Hausdorff distance, from 3.098 mm with a single encoder to 3.695 mm with the shared approach.
This shows that our model performs similarly in terms of overlap, but it is slightly less accurate in terms of boundary delineation.

 \begin{table}[tp]
 \centering
 \setlength\tabcolsep{0.25em}
    \caption{Abdominal plane segmentation results.}
     \label{tab:seg_abdomen}
     \begin{tabular}{lcc|cc}
         \cmidrule[1pt](l{0pt}r{0pt}){2-5}
         & \multicolumn{2}{c|}{\textbf{Dice Score }} & \multicolumn{2}{c}{\textbf{Hausdorff Distance (mm)}} \\ 
         \cmidrule(l{0pt}r{0pt}){2-5} 
         & \textbf{Median [IQR]}  & \textbf{5\%-worst} & \textbf{Median [IQR]}  & \textbf{5\%-worst} \\ 
         \midrule
         Individual Encoder  & 0.943 [0.933, 0.955]  & 0.912 & 8.210 [5.935, 10.295]  & 38.430  \\
         Shared Encoder      & 0.956 [0.938, 0.963]  & 0.872 & 6.736 [4.479, 14.093]  & 50.476  \\
         \bottomrule
     \end{tabular}
 \end{table}

 \begin{table}[tp]
     \centering
     \setlength\tabcolsep{0.25em}
    \caption{Head plane segmentation results.}
     \label{tab:seg_brain}
     \begin{tabular}{lcc|cc}
         \cmidrule(lr){2-5}
         & \multicolumn{2}{c|}{\textbf{Dice Score}} & \multicolumn{2}{c}{\textbf{Hausdorff Distance (mm)}} \\ 
         \cmidrule(lr){2-5} 
         & \textbf{Median [IQR]}  & \textbf{5\%-worst} & \textbf{Median [IQR]}  & \textbf{5\%-worst} \\ 
         \midrule
         Individual Encoder  & 0.974 [0.965, 0.980]  & 0.956 & 3.098 [2.503, 4.615]  & 6.151  \\
         Shared Encoder    & 0.973 [0.961, 0.978]  & 0.954 & 3.695 [2.892, 4.940]   & 6.217  \\
         \bottomrule
     \end{tabular}
 \end{table}

Finally, we assessed the accuracy of the biometric measurements.
Two ellipses were fitted to the extracted head and abdomen segmentation contours to calculate the perimeter lengths. Specifically, for the head plane, we utilised the minor axis of the fitted ellipse as an approximation of the biparietal diameter. 
For the femur, we applied a post-processing pipeline following the prediction of the distance map. This process involved the application of a Gaussian filter to reduce noise, and morphological operations to improve the separability of the two regions of interest. 
Then, the FL was calculated as the Euclidean distance between the minimum points in pixels. We converted all measurements to centimetres using the conversion factor specific to each image to apply the Hadlock equation for GA estimation. 
We assessed the performance of these measurements in terms of Mean Absolute Error (MAE), Mean Squared Error (MSE), Root Mean Squared Error (RMSE), and Mean Absolute Percentage Error (MAPE) to quantify the discrepancies between annotated and predicted values.
As shown in Table~\ref{tab:biometry_measures}, the MAE was less than 1 cm for 3 out of 4 measurements, only exceeding that value for the AC (1.207 cm).
With respect to the MAPE, none of the measurements exceeded the 6\%. This also includes GA, with a mean percentage error equal to 2.3\%, corresponding to a variation of ±0.76 weeks. 

\begin{table}[tp]
    \centering
    \setlength\tabcolsep{1em}
    \caption{Results of biometrical measures. HC: Head Circumference; BDP: Biparietal Diameter, AC: Abdominal Circumference, FL: Femur length, and GA: Gestational Age }
    \label{tab:biometry_measures}
    \begin{tabular}{lcccc}
        \cmidrule[1pt](l{0pt}r{0pt}){2-5}
        & \textbf{MAE} & \textbf{MSE} & \textbf{RMSE} & \textbf{MAPE} \\ 
        \midrule
        HC (cm)  & 0.658 & 0.576 & 0.759 & 0.021 \\
        BPD (cm) & 0.097 & 0.015 & 0.124 & 0.011 \\
        AC (cm)  & 1.207 & 2.790 & 1.670 & 0.038 \\
        FL (cm)  & 0.314 & 0.257 & 0.506 & 0.054 \\
        \midrule
        GA (weeks) & 0.762 & 1.091 & 1.045 & 0.023 \\
        \bottomrule
    \end{tabular}
\end{table}

\section{Discussion}

We have developed an automated GA estimation framework designed to achieve high accuracy despite the limited availability of training data. 
To address this challenge, we proposed a variant of the U-Net architecture that improves abdominal segmentation performance by combining U-Net blocks into a single-encoder.
This encoder exploits pre-training on structurally similar samples from the head plane, thereby improving generalisation.
The test results demonstrate that the proposed segmentation model performs as well as an independent U-Net explicitly trained for each task. 
These results highlight the ability of the model to reduce the total number of parameters required for segmentation by 25\% while maintaining statistically comparable performance to its independent counterpart. 
In addition, for FL computation, the method circumvents the lack of femur segmentation masks by relying solely on weak annotations (endpoints) used in fetal biometry as input data.

For the biometric model, the results are within the expected range of inter-operator variability, which is between 4.9\% and 11\% according to~\cite{sarris2012intra}. In particular, for 3  
of the measurements (HC, BPD, and AC) the error percentages are even lower than this range. 
Only the femur exceeds these values, with an average percentage error of 5.4\%.
However, it is reasonable to expect a higher error rate in the latter two cases (AC and FL). For AC, this may be due to the limited information available for segmentation and the use of a pre-trained model on an imbalanced dataset.
As a result, for the shared encoder approach, the Hausdorff distance is 1.8 times greater for abdominal segmentation than for head segmentation.
Furthermore, as illustrated in Fig.~\ref{Fig:Sistem_diagram}, the boundaries of the abdomen have lower contrast than the head, making accurate contour delineation more difficult. 
With respect to FL, the task is inherently complex due to the femur anatomical characteristics.  
The annotated endpoints are not placed at the bone boundaries but along the ossified diaphysis, which increases the inter-observer variability. In addition, as the femur is the smallest structure measured, even small variations can result in significant percentage discrepancies in the final measurement.

\section{Conclusions}

In this work, we have proposed an automated framework for GA estimation based on ultrasound imaging. 
Our results demonstrate that simpler architectures can achieve statistically similar performance to state-of-the-art methods, reducing the dependence on large datasets.
In addition, we found effective distance transformations for femur endpoint localisation, providing a reliable alternative for weakly annotated data. 
Together, these two models provide a framework for GA calculation with minimal variation, lower than the inter-operator variability reported for biometric measurements, highlighting its potential for clinical application, subject to further validation. 
Future work will focus on improving abdominal boundary delineation, and incorporating an optimal plane classification model to further automate GA estimation and reduce diagnostic time.

\footnotesize \section*{Acknowledgments}  
This work was supported by the Galician Ministry of Culture, Education, Professional Training and University (ED481A-2023-121, 2024-2027 ED431G2023/04, 2022-2025 ED431C2022/19, and GRC2021/48).
The previous grants are co-funded by the European Regional Development Fund (ERDF/FEDER program).
This work is also supported by the Spanish Ministry of Science and Innovation through the project 
COLLAGE PID2023-149959OA-I00; and by the Spanish Ministry of Science and Innovation, the Spanish Research Agency, and the European Union through the “Ramón y Cajal” 2022 program (RYC2022-035469-I and RYC2022-035960-I).

\bibliographystyle{splncs04}
\bibliography{mybibliography}

\end{document}